      [1999/12/01 v1.4c Il Nuovo Cimento]
\documentclass{cimento}
\usepackage{graphicx}

\title{A new approach of Analyzing GRB light curves}
\author{B.~Varga\from{ins:eot},
I.~Horv\'ath\from{ins:bol},
L.G.~Bal\'azs\from{ins:kon}}
\instlist{\inst{ins:eot} E\"otv\"os University, P\'azm\'any P\'eter s\'et\'any 1/A,
H-1518 Budapest, Hungary
  \inst{ins:bol} Dept. of Physics, Bolyai Military Univ. POB 12,
H-1456 Budapest, Hungary
  \inst{ins:kon} Konkoly Observatory, POB 67, H-1525 Budapest, Hungary}
\PACSes{\PACSit{98.70.-f}{98.80.-k}{~--~02.50.sk}}

\begin{document}

\maketitle

\begin{abstract}
We estimated the Txx quantiles of the cumulative GRB light curves
using our recalculated background. The basic information of the
light curves was extracted by multivariate statistical methods.
The possible classes of the light curves are also briefly
discussed.
\end{abstract}

\section{Introduction}

 The light curve represents one of the
most revealing information of a Gamma-ray Burst (GRB) (see, for
example, \cite{ref:me1}, \cite{ref:ry1}, \cite{ref:ry2} and the
references therein). In this work we make an attempt to classify
the bursts  by the light curves. We base this new approach on the
BATSE Concatenated 64-ms burst database \cite{ref:g1}, which
contains 2130 triggers. In order to be consistent with several
other papers (e.g. \cite{ref:bn}) we used the same background and
burst interval definitions as in \cite{ref:g2} but we made our own
independent parabolic background fits. We had background intervals
 for 2024 bursts \cite{ref:g2} and  we  made
 parabolic fits  using a $\chi^2$ \cite{ref:num} method
in the 1024 ms scale data, summed up the four channels. Then we
subtracted the estimated background from the BATSE cat64ms data
 and calculated $T_{90}$ (\cite{ref:bn},
\cite{ref:kos}, \cite{ref:mee}). Noisy bursts were excluded (where
the $\chi^2$ probability  was less than 0.0005), therefore the
final size of the sample was reduced to 1708. After fitting the
background we calculated the times when the 5\%, 10\%, . . . ,
95\% of the photons  were detected. For this purpose we calculated
the cumulative   light curves. After calculating the 19 time
values we computed the differences of the neighboring values
(t10\% - t5\%, t15\% - t10\%, t20\% - t15\%, etc.) and normalized
them with $T_{90}$, therefore the duration of the bursts became
uniformly one. This means that the duration information were
removed form the data.  These 18 variables were the base of our
analysis.

\section{Calculations}

\begin{figure}[ht]
  \begin{minipage}[t]{0.45\textwidth}
    \centering
    \includegraphics[scale=0.6]{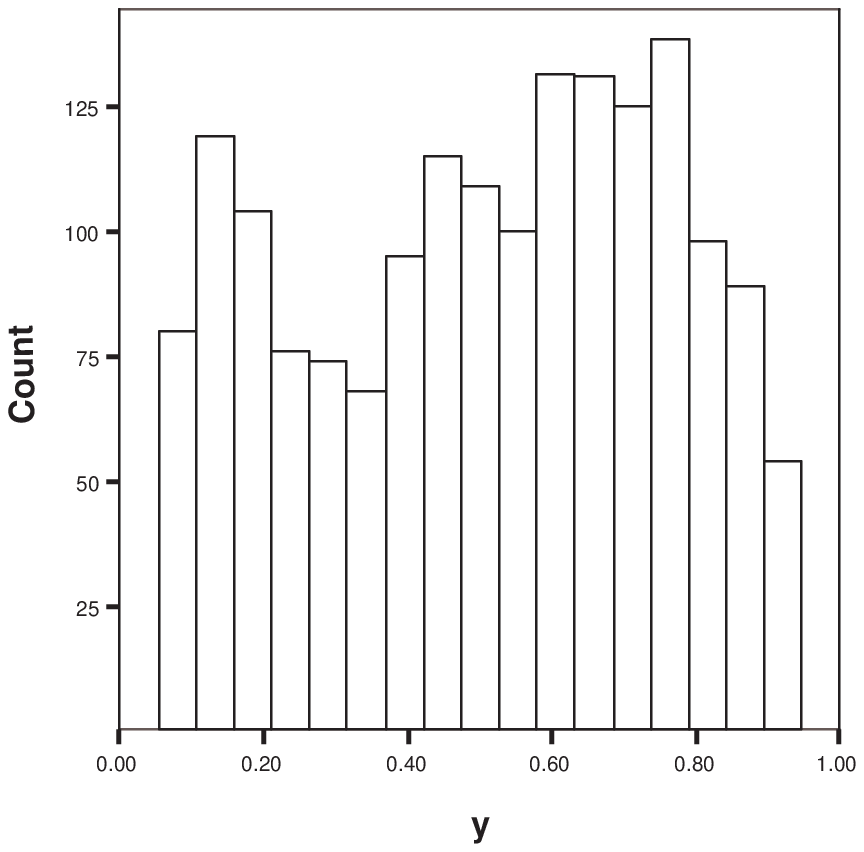}   
    \caption{\label{vcut} The histogram of the values where a
     vertical line near the maximal separation intersect
     the cumulative light curves.}
  \end{minipage}
  \begin{minipage}[t]{0.5\textwidth}
    \centering
    \includegraphics[scale=0.52, trim=0mm -12mm 0mm 0mm, clip]{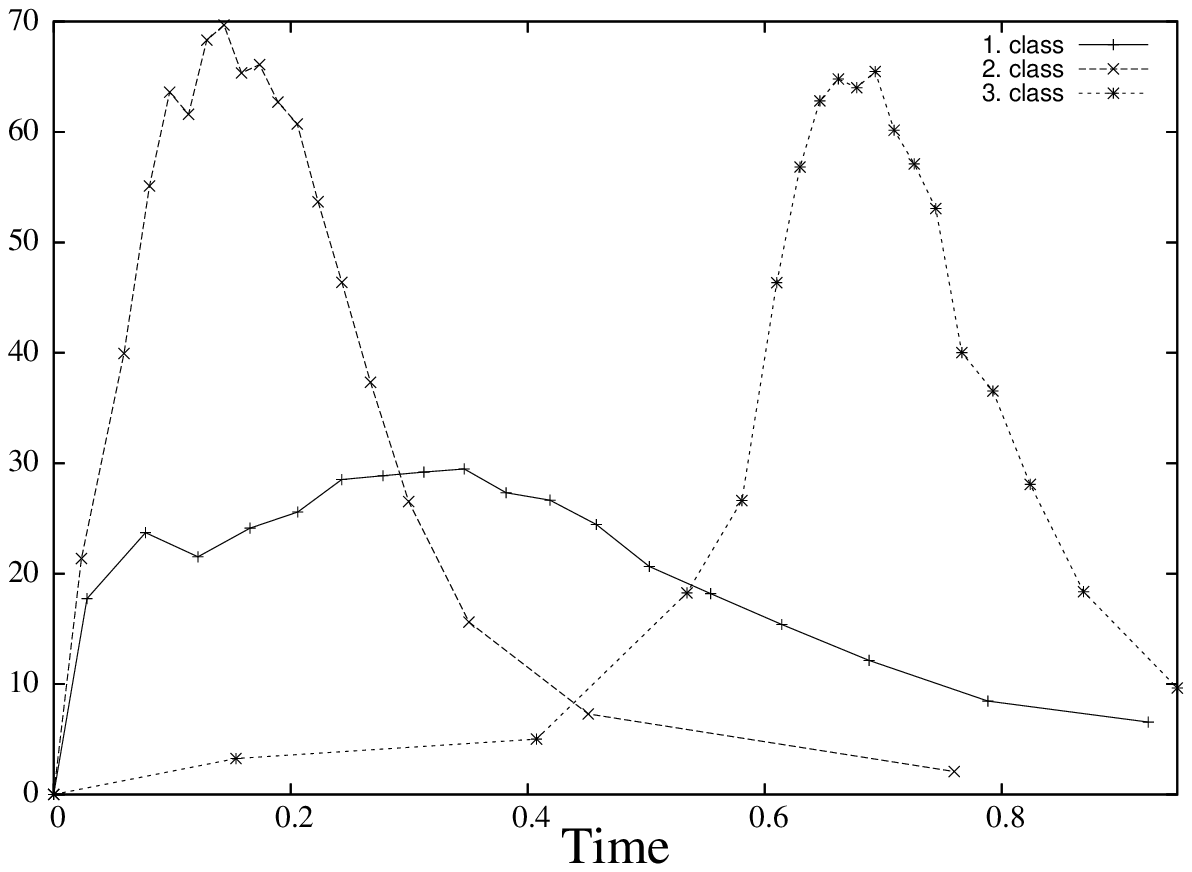}
    \caption{\label{lightc} The 3 light curves of the cluster centers.}
  \end{minipage}
    \
    \includegraphics[scale=0.52]{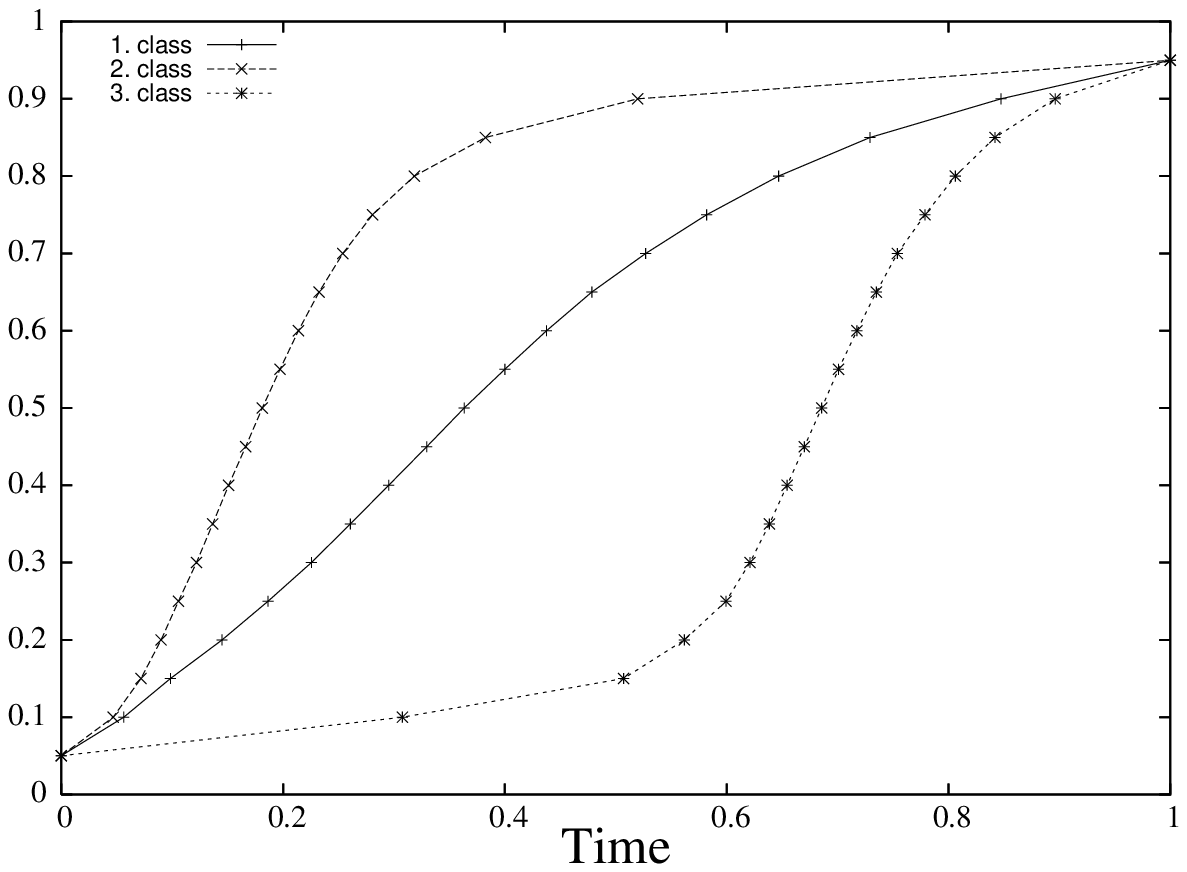}
    \includegraphics[scale=0.52]{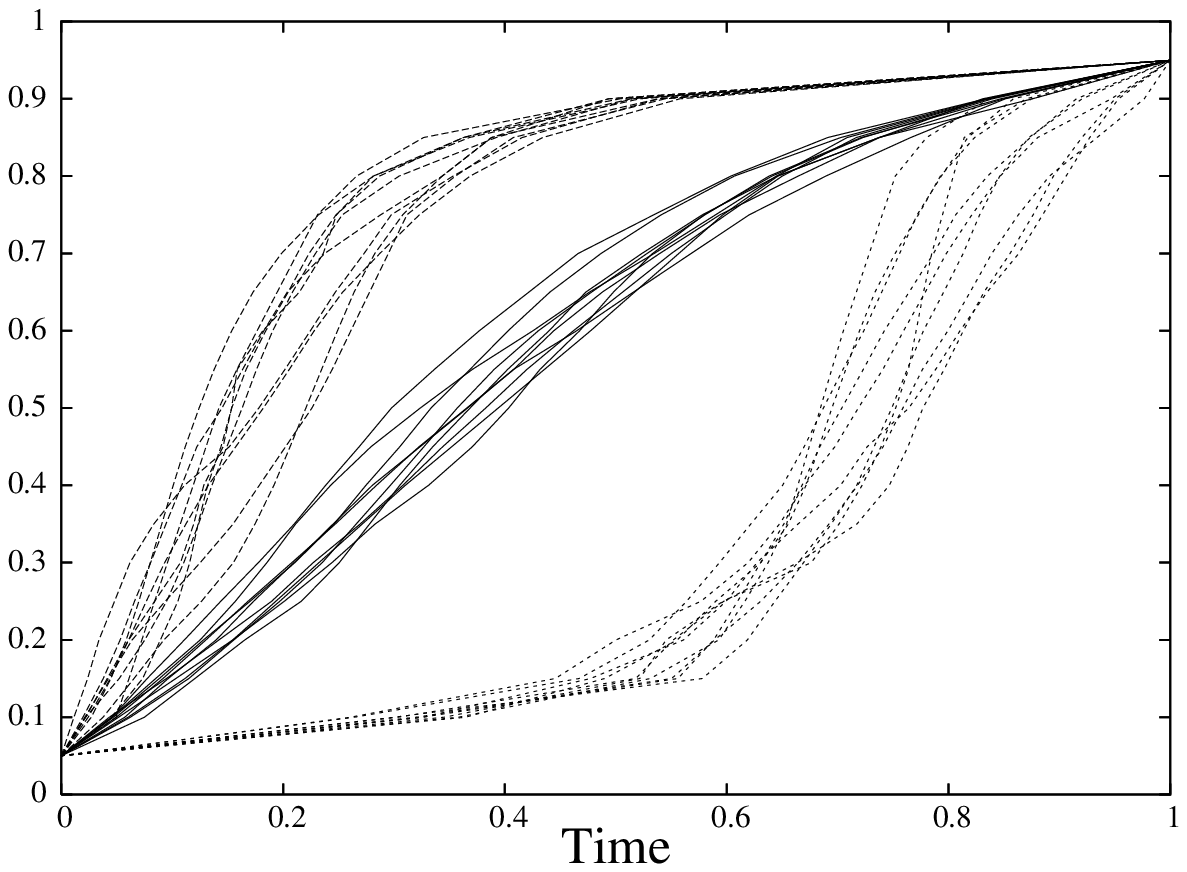}
    \caption{\label{cumu}On the left side we displayed the
     cumulative light curve  of the incoming photons, calculated from
     the cluster centers; on the right side 10 typical cumulative
     functions are selected from each class defined by the cluster
     analysis. (On the horizontal axis we displayed the relative time
     during the burst and on the vertical one the percentage of the
     detected photons.)}

\end{figure}

\begin{figure}[t]
  \includegraphics[scale=0.66]{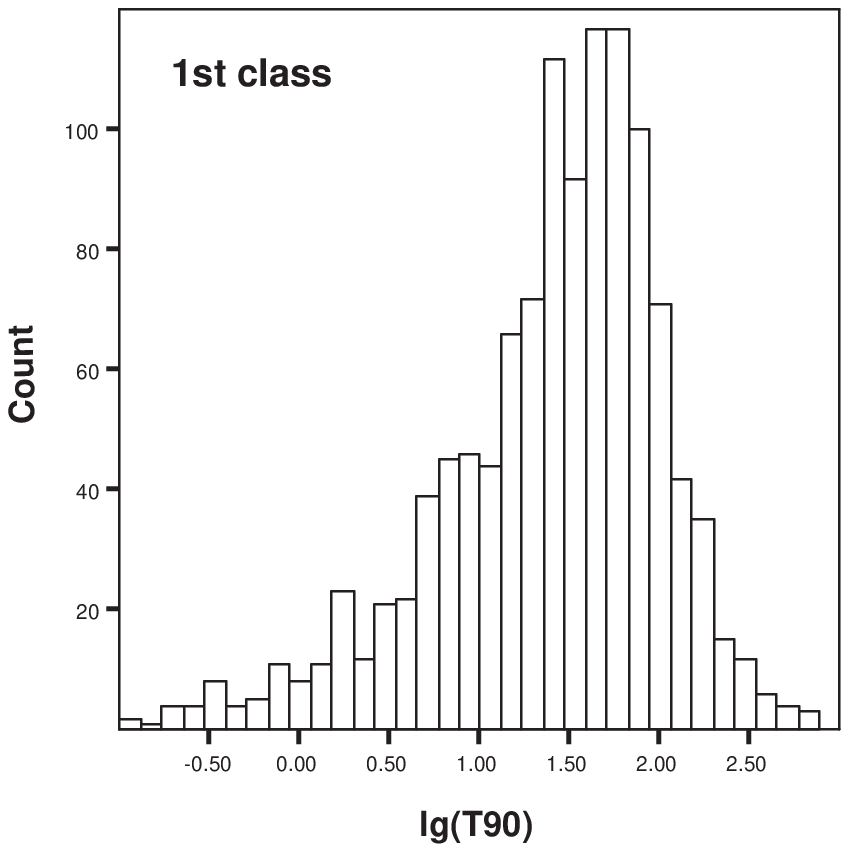}
  \includegraphics[scale=0.66]{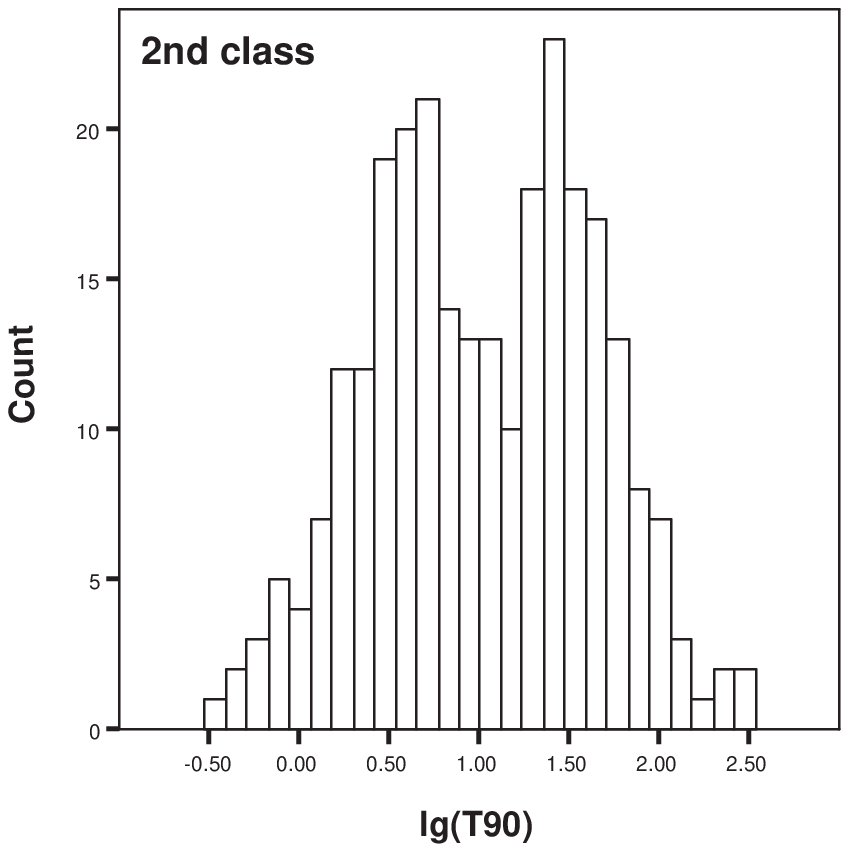}
  \includegraphics[scale=0.66]{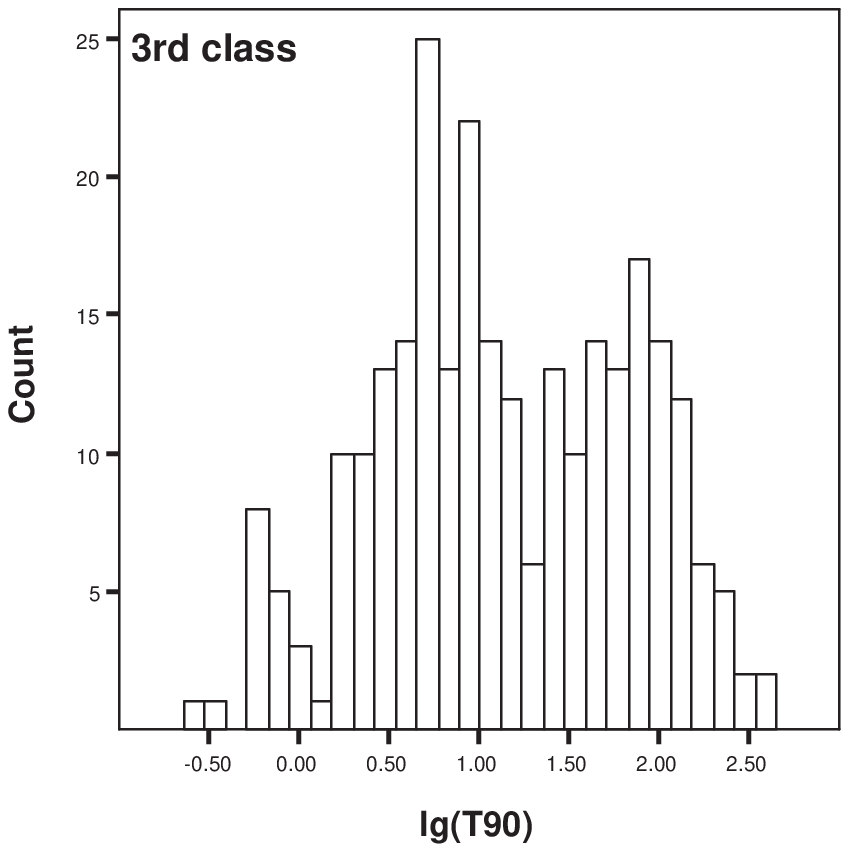}
  \includegraphics[scale=0.66]{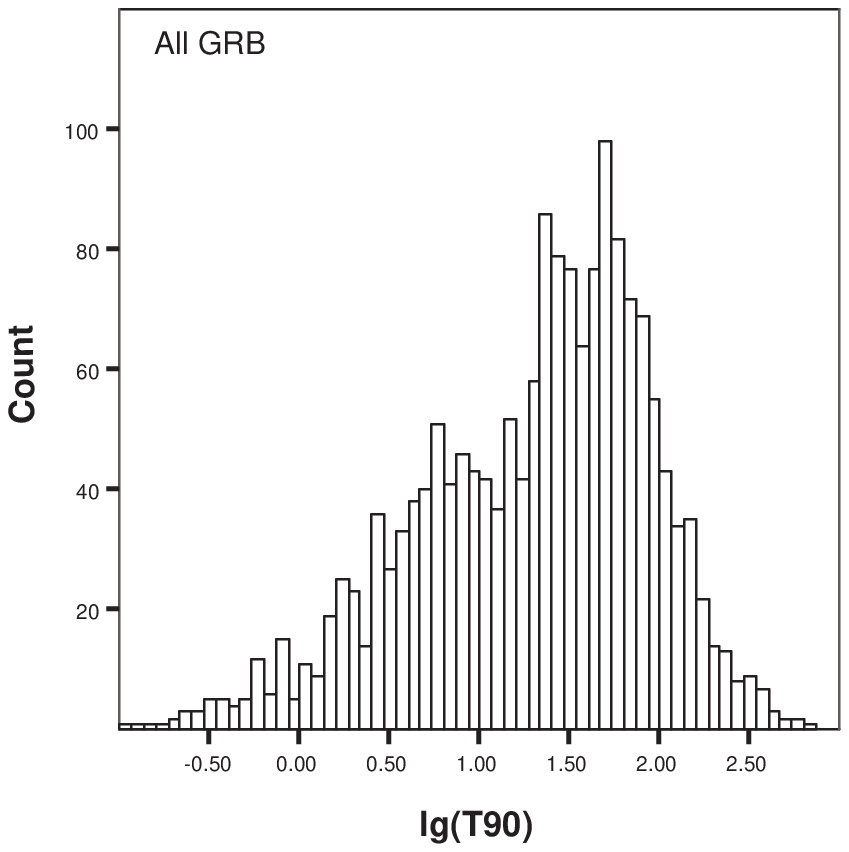}
  \caption{\label{his}Histograms of logarithmic T90  durations. 1st class
   on the top left, 2nd class on the top right, 3rd class on the bottom left and
   the all GRB on the bottom right corner.}
\end{figure}

We used the SPSS Statistical Analysis Software \cite{ref:sps} to
derive statistical information from our data set. For obtaining
the possible classes we used the {\it K}-means clustering procedure  
('Quick cluster' module in SPSS). This procedure assumes that data 
can be divided into a known {\it K} number of clusters (for GRB classification see
\cite{ref:bag}, \cite{ref:hak}, \cite{ref:ho1}, \cite{ref:ho2}, \cite{ref:mbv}
\cite{ref:muk}). For each 1708 bursts we had 18 variables, which
were the input for the {\it K}-means cluster analysis. The algorithm
tried to separate the bursts which were points in a 18 dimensional
space into the specified number of classes, therefore each burst
falls closer to its own class center than to  the other ones. The
distance was specified by the usual Euclidean distance. The $K$
value is an input parameter of the algorithm. After making a
vertical cut on the whole set of the cumulative light curves near the
maximal separation we recognized three clumps so we accepted $K=3$
(see Fig.~\ref{vcut}). This algorithm is sensitive to the initial setting
of cluster centers, so we started the iteration from several
different positions to make sure that the class centers obtained
are stable. The 1st group contained 1174, the 2nd group 268, and
3rd group 266 bursts. This procedure also gave the Euclidean
distances of each GRB from the center of the classes. In Fig.~\ref{cumu}
we displayed the cumulative light curves calculated from the
coordinates of the cluster centers, along with 10 typical (the
nearest ten to the center) cumulative curves from each classes.
From our data set we easily calculated the light curves that we
show in Fig.~\ref{lightc}. In this picture we can see that the classification
of the bursts in this 18 dimensional space happens according to the
distribution of arrival times of the photons. In the case of the
first class the photons were distributed almost smoothly in the whole
duration, in the second class most of the photons detected at the
beginning of the burst, and in the third one  at the end. After
assigning the bursts to classes we compared the real light curves
with  those of the cluster centers, and we found them to be
similar.

The next step was the analysis of the classes. We examined the
distribution of the logarithmic $T_{90}$ durations within each
class, and found that the distributions in 3 different classes
have different numbers of maxima, and have them at different
durations. The first class, which is the most populated, and has
got the smoothest cumulative function, has one maximum at 50s,
while the 2nd and 3rd groups have two maxima at 15s, 5s and 80s
and 5.6s (the histograms can be seen in Fig.~\ref{his}). This is not
obvious because the duration information was excluded from the
data set, by normalization. We also calculated the average
duration for each class. The first, second and third classes have
51.97s, 25.15s, 40.65s average duration, respectively. This means
that these three classes are  different in the mean duration.

All of the results are available on the World Wide Web \cite{ref:varg},
specially the class membership of the analyzed GRBs.

\section{Conclusion}
We calculated new Txx quantiles of the cumulative GRB light
curves. Making use the 'Quick cluster' module of the SPSS
statistical package we performed {\it K}-means clustering and identified
3 classes in our sample. The general form of the light curve
differs between the classes. Comparing
the frequency distribution of the newly determined $T_{90}$
durations we recognized that the diagram of the 1st class peaks at
long $T_{90}$ value, the 2nd has two and the 3rd maybe three
maxima. These new results may give a new insight into the physical
classification of GRBs.

\acknowledgments

 This research was supported through OTKA grants T034549
and T48870.

\end{document}